\documentclass[aps,prb,twocolumn,longbibliography,showpacs,superscriptaddress,floatfix]{revtex4-2}
\usepackage{graphicx}
\usepackage{amsmath,amssymb,bbold,bm,color}
\usepackage{float}
\usepackage{epstopdf}
\usepackage{hyperref}
\usepackage{color}
\usepackage{enumerate}
\usepackage{wasysym}
\usepackage{url}%Turn on when output to pdf file
\usepackage{dsfont}
\usepackage{braket}
\usepackage{comment}
\hypersetup{colorlinks=true,linkcolor=blue,citecolor=blue,urlcolor=blue}

\makeatletter
\def\mathcolor#1#{\@mathcolor{#1}}
\def\@mathcolor#1#2#3{%
	\protect\leavevmode
	\begingroup
	\color#1{#2}#3%
	\endgroup
}
\makeatother

\begin{document}

\title{Traversable wormhole in coupled SYK models with imbalanced interactions}

\author{Rafael Haenel}
\affiliation{Department of Physics and Astronomy \& Stewart Blusson Quantum Matter Institute, University of British Columbia, Vancouver BC, Canada V6T 1Z4.}
\affiliation{Max Planck Institute for Solid State Research,
	70569 Stuttgart, Germany}
\author{Sharmistha Sahoo}
\affiliation{Department of Physics and Astronomy \& Stewart Blusson Quantum Matter Institute, University of British Columbia, Vancouver BC, Canada V6T 1Z4.}
\author{Timothy H. Hsieh}
\affiliation{Perimeter Institute for Theoretical Physics, Waterloo, Ontario N2L 2Y5, Canada}
\author{Marcel Franz}
\affiliation{Department of Physics and Astronomy \& Stewart Blusson Quantum Matter Institute, University of British Columbia, Vancouver BC, Canada V6T 1Z4.}

\date{\today}

\begin{abstract}
  A pair of identical Sachdev-Ye-Kitaev (SYK) models with bilinear coupling forms a quantum dual of a traversable wormhole, with a ground state close to the so called  thermofield double state. We use adiabaticity arguments and numerical simulations  to show that the character of the ground state remains unchanged when interaction strengths in the two SYK models become imbalanced. Analysis of thermodynamic and dynamical quantities highlights the key signatures of the wormhole phase in the imbalanced case. Further adiabatic evolution naturally leads to the `maximally imbalanced'  limit where fermions in a single SYK model are each coupled to a free Majorana zero mode. This limit is interesting because it could be more easily realized using various setups proposed to implement the SYK model in a laboratory. We find that this limiting case is marginal in that it retains some of the characteristic signatures of the wormhole physics, such as the spectral gap and revival dynamics,  but lacks others so that it likely does not represent the full-fledged wormhole dual. We discuss how this scenario could be implemented in the proposed realization of the SYK physics in quantum wires of finite length coupled to a disordered quantum dot.
\end{abstract}

\maketitle

\section{Introduction}
The proposal of the SYK model~\cite{SY1993,Kitaev2015,MaldacenaStanford2016} as dual holographic description for a quantum black hole in (1+1) dimensional anti-de Sitter space AdS$_{2}$ has motivated a flurry of research in scrambling of quantum information and many-body quantum chaos. The model consists of $N$ Majorana fermions interacting through random all-to-all interactions and is defined by the Hamiltonian
\begin{equation}
	\label{eqn:SYK}
	H_{\mathrm{SYK}} = \sum_{i< j< k < l}^{}
	J_{ijkl}\chi_{i}\chi_{j}\chi_{k}\chi_{l}.
\end{equation}
Here, $\chi_{i}$ are Majorana fermion operators that satisfy the anti-commutation relations $\left\{\chi_{i}, \chi_{j} \right\}=\delta_{ij}$ and $J_{ijkl}$ are real constants drawn from a Gaussian ensemble obeying 
$\langle J_{ijkl}^2 \rangle =  {6} J^2/{N^3}$.

Two {\em identical} SYK models coupled via one-to-one bilinear coupling have been shown to realize a holographic dual of an eternal traversable wormhole \cite{maldacenaqi2018}.
The quantum ground state corresponding to the wormhole is a thermofield double (TFD) state that can be regarded as a  particular purification of the thermal Gibbs ensemble at inverse temperature $\tilde{\beta}$ and is given by
\begin{equation}
\ket{ \rm{TFD}_{\tilde\beta}} = \frac{1}{\sqrt{Z_{\tilde\beta}}} \sum_n e^{-{\tilde\beta} E_n/2} \ket{n}_L \otimes { \ket{\Theta n}}_R \,.
\label{eqn:TFD}
\end{equation}
Here, $\ket{n}_L$ and $\ket{n}_R$ are eigenstates of two copies of the original system, denoted as `left' and `right'. They are related by an anti-unitary symmetry $\Theta$ such that $\ket{n}$ and $\Theta \ket{n}$ are degenerate with energy $E_n$, and $Z_{\tilde{\beta}}=\sum_n e^{-\tilde{\beta}E_n}$ is the partition function of a single system.

Preparing the TFD state for an arbitrary $\tilde \beta$ is a non-trivial task, which was made possible in the Ising model through a variational method \cite{Hsieh2019, Zhu2020}. In a parallel series of studies it was shown that TFD state can occur as an exact or approximate ground state of two identical copies of a system, provided they are coupled in a specific way \cite{maldacenaqi2018,Cottrell2019,sahoo2020}. Here, the coupling strength sets the temperature $\tilde{\beta}$ of the TFD. For example, in Ref.~\onlinecite{maldacenaqi2018} by Maldacena and Qi (MQ), the ground state of two identical SYK Hamiltonians \eqref{eqn:SYK} coupled by
\begin{equation}
H^{LR}=i\mu\sum_{j}^{}\chi^L_j \chi^R_j
\label{hlr}
\end{equation}
has been shown to satisfy
\begin{equation}
	\left( H^L_{\mathrm{SYK}} + H^R_{\mathrm{SYK}} + H^{LR}\right) \ket{\mathrm{TFD}} \approx E_{GS}
	\ket{\mathrm{TFD}}
	\label{eqn:coupled_SYK}
      \end{equation}
with high fidelity. In the gravity context, such coupling leads to negative null energy making the wormhole \emph{traversable}, i.e., particles can be sent through it \cite{Yang2017,Bak2018,Maldacena2018,Gao2017,Gao2019,Gao2019b}. The MQ model exhibits sharp revivals of scrambled local perturbations \cite{Plugge2020} and undergoes a finite temperature Hawking-Page transition \cite{Hawking1983} between wormhole and black hole phases \cite{Garcia-Garcia2019}.  These effects  have been identified as characteristic signatures of the wormhole phase.

Significant efforts have been made to physically realize  the SYK model in atomic~\cite{Danshita2017}, optical~\cite{wei2020optical}, and solid-state~\cite{Pikulin2017,Alicea2017,Rozali2018} platforms, or using quantum simulators~\cite{Solano2017,Luo2019}. Proposals for  implementing and observing the wormhole physics in the MQ model and its complex fermion variants have also been made \cite{LH2019,sahoo2020,zhou2020}.  The key challenge in these proposals is that one must realize two SYK systems with random couplings $J_{ijkl}$ that are {\em identical} at the microscopic level. In solid-state realizations such randomness inevitably comes from disorder which cannot be controlled, hence the challenge. 

Here, we present an approach to this challenge based on a simple adiabatic argument. We consider a variant of the MQ model with identical interactions in $L$ and $R$ subsystems except for an overall scale factor, leading to an MQ model with imbalanced interaction strengths. We show that wormhole behavior persists in this imbalanced model all the way to the limit when the interaction strength in one of the systems vanishes. This maximally imbalanced limit represents an interesting  marginal case which, as we show, still  exhibits some vestiges of the wormhole physics.  Crucially, this case could be amenable to experimental study because it only  requires realization of  a {\em single} SYK system whose constituent fermions are each coupled  to a free Majorana mode.
\begin{figure}
	\centering
	\includegraphics[width=0.5\textwidth]{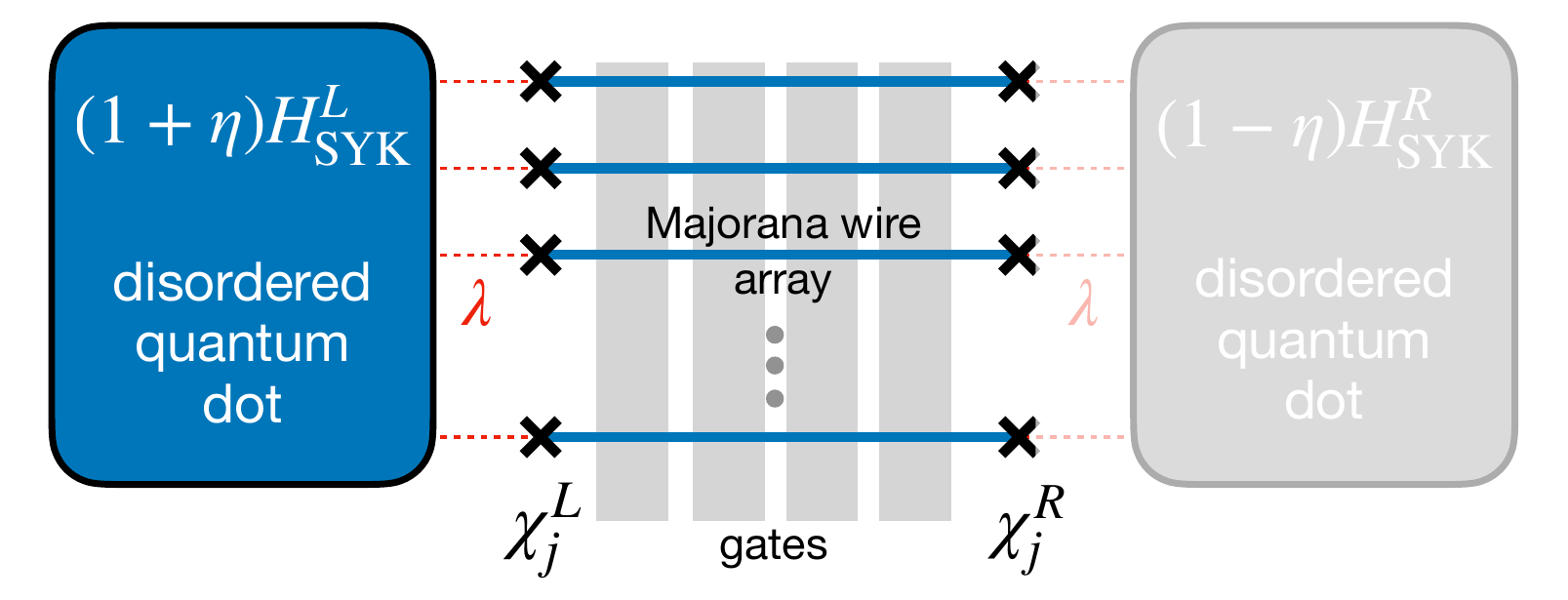}
	\caption{A possible realization of the imbalanced Maldacena-Qi model based on the SYK setup proposed in  Ref.~\onlinecite{Alicea2017}. Majorana zero modes localized at the two ends of proximitized quantum wires are weakly coupled to two disordered quantum dots, $L$ and $R$. We consider relatively short wires such that $\chi_j^L$ and $\chi_j^R$ are weakly hybridized due to the overlap between their wavefunctions inside each wire.  Because microscopic realizations of disorder will inevitably be different in $L$ and $R$ quantum  dots the corresponding $J^L_{ijkl}$ and $J^R_{ijkl}$ will be uncorrelated. However, when the $R$ dot is removed altogether then the setup trivially realizes the maximally imbalanced MQ model defined by Eq.~\eqref{heta2} with $\eta=1$. }
	\label{fig0}
\end{figure}

The idea relies on the identity
\begin{equation}
\left(H^L_{\mathrm{SYK}}-H^R_{\mathrm{SYK}}\right)\ket{\mathrm{TFD}} = 0
\label{iden}
\end{equation}
which can be verified directly from the definition of the TFD state, Eq.~(\ref{eqn:TFD}). Combining this with Eq.~\eqref{eqn:coupled_SYK} it follows that  $\ket{\mathrm{TFD}}$ also satisfies
\begin{equation}\label{heta1} 
H_\eta\ket{\mathrm{TFD}} 
\approx 
E_{\rm GS}
\ket{\mathrm{TFD}},
\end{equation}
with the Hamiltonian
\begin{equation}\label{heta2}
H_\eta=(1+\eta)H^L_{\mathrm{SYK}} + (1-\eta)H^R_{\mathrm{SYK}} + H^{LR}
\end{equation}
which defines the imbalanced MQ model. If $\ket{\mathrm{TFD}}$ was an {\em exact} ground state of the original MQ Hamiltonian $H_{\eta=0}$ then by virtue of identity \eqref{iden} it would remain an exact eigenstate of $H_\eta$ for any value of $\eta$. Because, however, Eq.~\eqref{eqn:coupled_SYK} holds only approximately, it remains to be seen how well Eq.~\eqref{heta1} holds away from $\eta=0$. On general grounds we expect that when $\eta$ is slowly increased from 0, $\ket{\rm TFD}$ remains the ground state as long as the gap in the spectrum does not close. We also note that, as pointed out in Ref.~\onlinecite{Hsieh2016}, for noninteracting fermions (i.e.\ systems described by $H^{L,R}$ quadratic in fermion operators) the  ground state of $H_{\eta=1}$ is {\em exactly} the same as the ground state of  $H_{\eta=0}$.

In the rest of this work we show, through a combination of exact diagonalization (ED) and large-$N$ saddle point solutions, that the system defined by Hamiltonian \eqref{heta2} indeed remains gapped for $\eta$ between 0 and $\eta_c$ at any finite $\mu$. The critical imbalance parameter $\eta_c>1$ depends on $\mu$ and approaches 1 from above as $\mu\to 0$. At the special point $\eta=1$, the TFD state can be realized approximately as the ground state of 
$2 H^L_{SYK} +H^{LR}$. As mentioned above this construction requires only a single realization of the SYK model and therefore circumvents the enormous challenge of fabricating two SYK models with identical couplings $J_{ijkl}$. We will discuss below how a small modification to the setup proposed in Ref.~\onlinecite{Alicea2017} to realize SYK physics in Majorana wires coupled to a quantum dot could be used to implement and probe the wormhole physics in the maximally imbalanced $\eta=1$ limit. The proposed setup is sketched in Fig.~\ref{fig0}.

Previously, three key signatures of the wormhole phase have been identified in the literature: (i) the presence of the thermofield-double (TFD) ground state, (ii) a temperature-driven first order phase transition of Hawking-Page type, and (iii) revival dynamics showing transmission of excitations between two maximally chaotic subsystems. We will investigate these indicators and find that all three  persist in the imbalanced MQ model for $\eta<1$. At $\eta=1$ the Hawking-Page transition disappears and a \emph{weak} form of the revival dynamics is observed. These observations suggest that the imbalanced MQ model is dual to a traversable wormhole for all $\eta<1$.  The maximally imbalanced case $\eta=1$ shows some tantalizing signatures of the wormhole physics but is likely not a full-fledged wormhole dual.

In Sections \ref{subsec:TFDgdstate} and \ref{subsec:Phasediagram} we present an analysis of the ground state and the phase diagram for general values of the tunneling strength $\mu$ and the imbalance parameter $\eta$. We discuss the revival dynamics of the ground state for the imbalanced case in Sec.~\ref{subsec:Revivaldynamics}. In Sec.~\ref{sec:eta1limit}, we analyze the revival dynamics for the $\eta=1$ limit and  discuss how it differs from the characteristics revival dynamics of a canonical MQ wormhole. The realization of such a limit in the proposed SYK model implementation of Ref.~\onlinecite{Alicea2017}  is also discussed here. We conclude with a summary and  open questions in Sec.~\ref{sec:conclusion}.

\section{The model and its properties\label{sec:modelgeneraleta}}
\begin{figure*}
	\begin{center}
		\includegraphics[width=\textwidth]{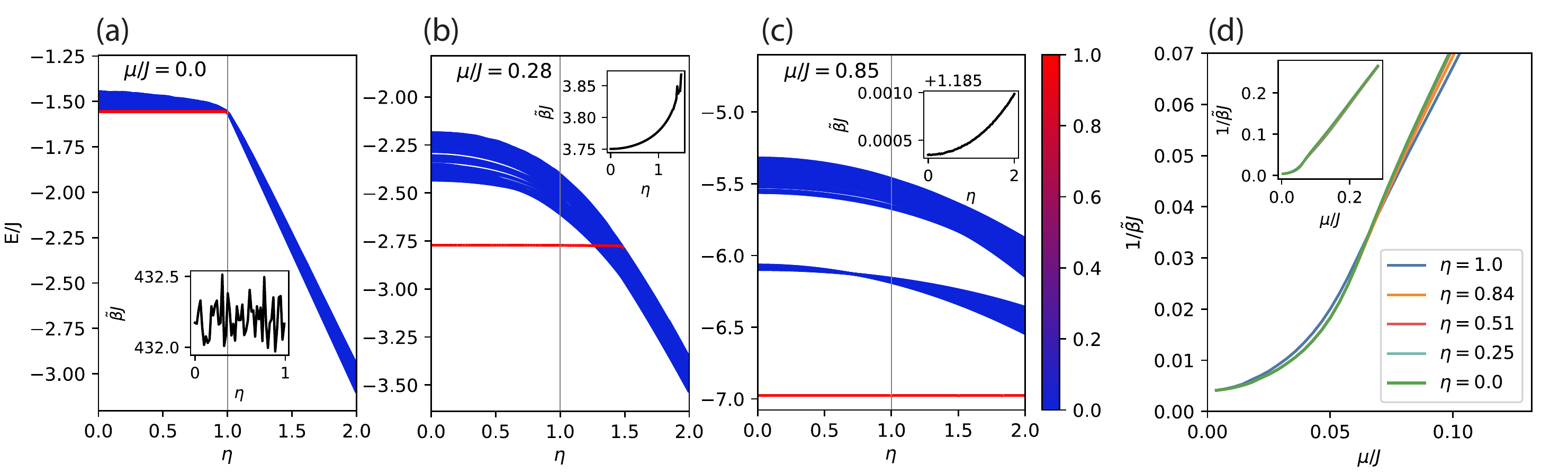}
	\end{center}
	\caption{(a-c) Spectrum of $50$ lowest energy levels of the imbalanced MQ Hamiltonian \eqref{heta2} obtained from ED for $2N=32$ for a single realization of Gaussian random variables $J_{ijkl}$. Colorscale indicates fidelity $f_{\rm TFD}(E_n)$  of the corresponding eigenstate with the TFD state, optimized with respect to the inverse temperature $\tilde{\beta}$, as explained in the main text. Insets show a weak dependece of $\tilde{\beta}$ on $\eta$ for $\mu>0$. (d) Plot of the inverse TFD temperature $\tilde{\beta}$ that maximizes the ground state fidelity $f_{\rm TFD}(E_0)$ as a function of $\mu$. Inset shows the linear relationship between $1/\tilde{\beta}$ and $\mu$ at large $\mu$, with only a week dependence on $\eta$.}
	\label{fig:ed}
\end{figure*}

The Hamiltonian of the imbalanced MQ model can be written  more compactly as
\begin{equation}
\begin{aligned}
\label{modelH}
H_\eta &= H^L + H^R + H^{LR} \\&= \sum_{\alpha=L,R}^{}(1+\alpha \eta)\sum_{i< j< k < l}^{}
J_{ijkl}\chi_{i}^{\alpha}\chi_{j}^\alpha\chi_{k}^\alpha\chi_{l}^\alpha
\\
&+  \frac{i\mu}{2} \sum_{j}^{} \left(\chi_j^L \chi_j^R - \chi_j^R
\chi_j^L \right) \,
\end{aligned}
\end{equation}
where index $\alpha$ assumes values $(+1,-1)$ for $(L,R)$ in numerical expressions.   
A constraint on the coupling, $J_{ijkl} = -J_{jikl} = -J_{ijlk} = J_{jilk}$, is imposed since fermion operators anticommute. 
Unlike the canonical MQ model  Hamiltonian \eqref{modelH}  lacks the discrete mirror symmetry defined by  $\chi_{j}^{L} \rightarrow \chi_{j}^{R}, \quad \chi_{j}^{R} \rightarrow -\chi_{j}^{L}$. The balanced limit ($\eta=0$) also has an additional global $\mathbb{Z}_{4}$ ``spin symmetry" which the imbalanced model lacks. If the spin operator is defined as $S=i \sum_{i=1}^{N} \chi_{i}^{L}\chi_{i}^{R}$ such that $H^{LR} = \mu S$, then for $\eta=0$ it holds that
\begin{align} 
\left[e^{i\pi S}, H_{\eta=0}\right] = 0, \quad \left[e^{i \pi S/2}, H_{\eta=0}\right] = 0.
\end{align}
Therefore $S$ mod $2$, i.e.\ chirality, and $S$ mod $4$ are both symmetries of the MQ Hamiltonian. When  $\eta \neq 0$ we are left with only chirality as a symmetry. These differences have implications for the level statistics analysis \cite{Garcia-Garcia2019} and for the ground-state degeneracy.  For any $\eta$ the model has antiunitary symmetry $\Theta$ that transforms $\chi_{L} \rightarrow \chi_{L}$, $\chi_{R} \rightarrow -\chi_{R}$ along with complex conjugation. This symmetry is crucial for our purposes as it enters the definition of the TFD state in Eq.~\eqref{eqn:TFD}.

\subsection{TFD ground state\label{subsec:TFDgdstate}}

We use a spinor representation ($2^{N}$ dimensional matrix) of Majorana fermion $\chi_{i}^\alpha$ operators satisfying the anticommutation algebra $\left\{\chi_{i}^\alpha, \chi_{j}^\beta \right\}=\delta_{ij}\delta^{\alpha\beta}$ to perform an ED study. Here $N$ denotes the number of Majorana fermions in one subsystem for the total of $2N$ Majoranas. The degeneracy of the SYK ground state depends on $N$ with a unique SYK ground state guaranteed when $N$ is a multiple of $8$. Here we focus on systems with $2N=16$ and $32$ to numerically find the lowest $50$ eigenvalues $E_n$ and corresponding eigenvectors $\ket{E_n}$. We determine the TFD inverse temperature ${\tilde{\beta}}$ by minimizing the ground-state fidelity $f_{\rm TFD}(E_0) = |\braket{E_0|{\rm TFD}_{\tilde{\beta}}}|^2$  with respect to $\tilde{\beta}$.
For this TFD temperature we then calculate $f_{\rm TFD}(E_n)$ for the excited states.
While the TFD inverse  temperature $\tilde{\beta}$ of the ground state depends on the coupling strength $\mu$, we find $\tilde{\beta}$ to be nearly independent of $\eta$.

Fig.~\ref{fig:ed}(a-c) shows the evolution of the $50$ lowest energy levels $E_n$ as a function of $0 \le \eta \le 2$ for several different values of coupling strength $\mu$. The colorscale indicates the overlap $f_{\rm TFD}(E_n)$. 
In all three panels we clearly observe a single state (highlighted in red) with the overlap close to 1.
As expected from our adiabatic argument, the energy of that state is nearly unaffected by $\eta$, at least for the small number of Majoranas $2N=16$ and $32$ considered in our ED study. 
In the first panel ($\mu=0$) the two SYK models are uncoupled and hence the eigenstates are of the form $\ket{n} \otimes \ket{m}$ with energies $E_{n,m}=E_n+E_m+\eta(E_n-E_m)$. 
In the large-$N$ limit, the model is gapless, however in the ED calculation, the model has a finite-size gap that is exponentially small in $N$. 
This allows for observation of the TFD state even at $\mu=0$ which would otherwise 
be hidden in the degenerate ground state manifold.
\begin{figure*}
	\centering
	\includegraphics[width=0.8\textwidth]{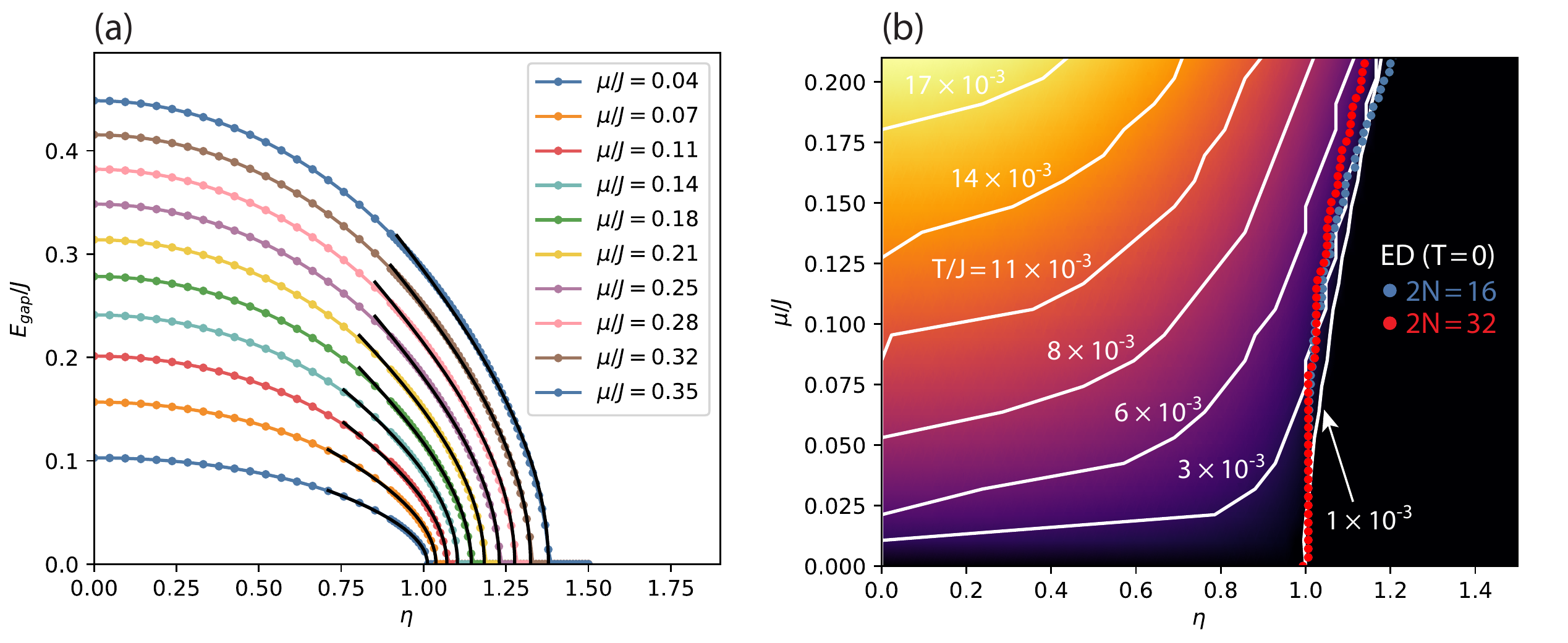}
	\caption{(a) Gap from the saddle point solutions as function of $\eta$ for various $\mu$ values indicating existence of a small gap even when $\eta=1$ only if $\mu$ is non-zero. (b) Phase boundary (whites lines) between gapped and gapless phase for various temperature values, $T$ (black lines). On the left of the line the ground state is gapped whereas it is gapless on the right. The lines end at a critical point. Blue and red dots trace phase boundary as obtained by exact diagonalization, averaged over $10$ realizations of Gaussian random couplings $J_{ijkl}$. Colors indicate the gap value computed from the imaginary time saddle point equations at $T/J=0.001$ where the gapless region is colored black.}
	\label{fig:largeNgap}
      \end{figure*}

For $\mu>0$ the energy level dependence on $\eta$ becomes nonlinear. The gap between the ground state and the excited states decreases as $\eta$ grows and eventually closes for critical strength of $\eta$. In the limit of $\mu \rightarrow 0$, the gap closes exactly at the special point $\eta=1$. For finite $\mu$, the critical point shifts to larger $\eta>1$. Below we compare this with the phase diagram in the parameter space of $\mu$ and $\eta$. Fig.~\ref{fig:ed}(d) shows $\tilde{\beta}$ corresponding to the Gibbs distribution in the TFD definition and indicates that it is a monotonic function of $\mu$. Note that the functional form of $\beta(\mu)$ is only weakly dependent on $\eta$. The dependence of the ground state on $N$ is discussed in Appendix~\ref{Appendix:tfdstate} and it is shown that as $N$ increases, $(H^{L}-H^{R})_{\eta=0}\ket{GS} \neq 0$ as the ground state is only \emph{close} to a TFD.

\subsection{Phase diagram\label{subsec:Phasediagram}} 
We supplement the above discussion based on ED for small number $N$ of Majorana fermions with a saddle-point solution that becomes exact in the large $N$ limit, $\beta J \ll N$. Standard procedure (outlined in Appendix~\ref{apdx:sd})
     % that involves reformulating the model as an imaginary-time path integral defining the  partition function and averaging over quenched disorder using the replica formalism,
      leads to a system of saddle-point equations  for the time-ordered fermion correlator $G_{\alpha \beta}(\tau, \tau') =\frac{1}{N}\langle \mathcal{T} \sum_{i=1}^{N} \chi^{\alpha}_{i}(\tau) \chi^{\beta}_{i}(\tau') \rangle$ and the associated self energy $\Sigma_{\alpha \beta}(\tau, \tau')$ which, for the imbalanced MQ model, take the form
\begin{equation}
\begin{aligned}
\label{eqn:sd}
G_{\alpha \alpha}(i\omega_n) &=
\frac{i\omega_n -\Sigma_{\bar{\alpha}\bar{\alpha}(i\omega_n)}}{D(i\omega_n)},
\\
G_{LR}(i\omega_n) &= 
-\frac{i\mu -\Sigma_{LR}(i\omega_n)}{D(i\omega_n)},
\\
\Sigma_{\alpha\beta}(\tau-\tau') &= J^2 (1+\alpha\eta)(1+\beta\eta) G_{\alpha\beta}(\tau-\tau')^3,
\end{aligned}
\end{equation}
where $\bar{\alpha}=-\alpha$ and
\begin{equation*}
D(i\omega_n) = 
\left(i\omega_n - \Sigma_{RR}\right)
\left(i\omega_n - \Sigma_{LL}\right)
+
\left(i\mu - \Sigma_{LR}\right)^2.
\end{equation*}
We used the anti-symmetry property of the fermion fields to impose $G_{\alpha \beta}(\tau) = - G_{\beta \alpha}(-\tau)$ and thus reduce the number of equations. 

For $\mu=0$ these equations can be \emph{exactly} solved in the low frequency limit ($\beta J \gg \omega$), giving a conformal solution for the diagonal Green's functions, $G_{\alpha \alpha}(\tau) = b \tau^{-\frac{1}{2}}$, with $b=1/(2\pi J^2 (1+\alpha \eta))^{\frac{1}{4}}$, while the off-diagonal Green's function, $G_{LR}(\tau)$ vanishes. From the numerical solution of Eq.~\eqref{eqn:sd} [see Fig.~\ref{fig:largeNgap}(a) and discussion below], we find that for $\eta \le 1$, an infinitesimal $\mu$ gaps out the model, while for $\eta>1$, the model remains gapless until a critical value for $\mu$. In the parameter range where the model is gapless, the low frequency conformal solution has a scaling symmetry in $\eta$. Given a solution $\left\{G_{\alpha \beta}, \Sigma_{\alpha \beta}\right\}$ at $(\eta, \mu)$, we can construct a new solution $\{\tilde{G}_{\alpha \beta}, \tilde{\Sigma}_{\alpha \beta}\}$ at $(\tilde{\eta}, \tilde{\mu} =  \frac{1-\tilde{\eta}^2}{1-\eta^2}\, \mu)$ by rescaling the correlators according to 
$$\tilde{G}_{\alpha \beta}=\text{sign}(c_{\alpha \beta})\left|c_{\alpha \beta}\right|^{1/4} G_{\alpha \beta},\ \ 
\tilde{\Sigma}_{\alpha \beta}=\left|c_{\alpha \beta}\right|^{-1/4} \Sigma_{\alpha \beta},$$ where 
\begin{equation}
  c_{\alpha \beta}=\frac{(1+\alpha \eta)(1+\beta \eta)}{(1+\alpha \tilde{\eta})(1+\beta \tilde{\eta})}.
\end{equation}
We find that the conformal solutions at different points in the gapless region of the $(\eta,\mu)$ phase diagram are related by an overall scale factor and decay with the SYK power law $G \propto |\tau|^{-1/2}$.

\begin{figure*}
	\centering
	\includegraphics[width=\textwidth]{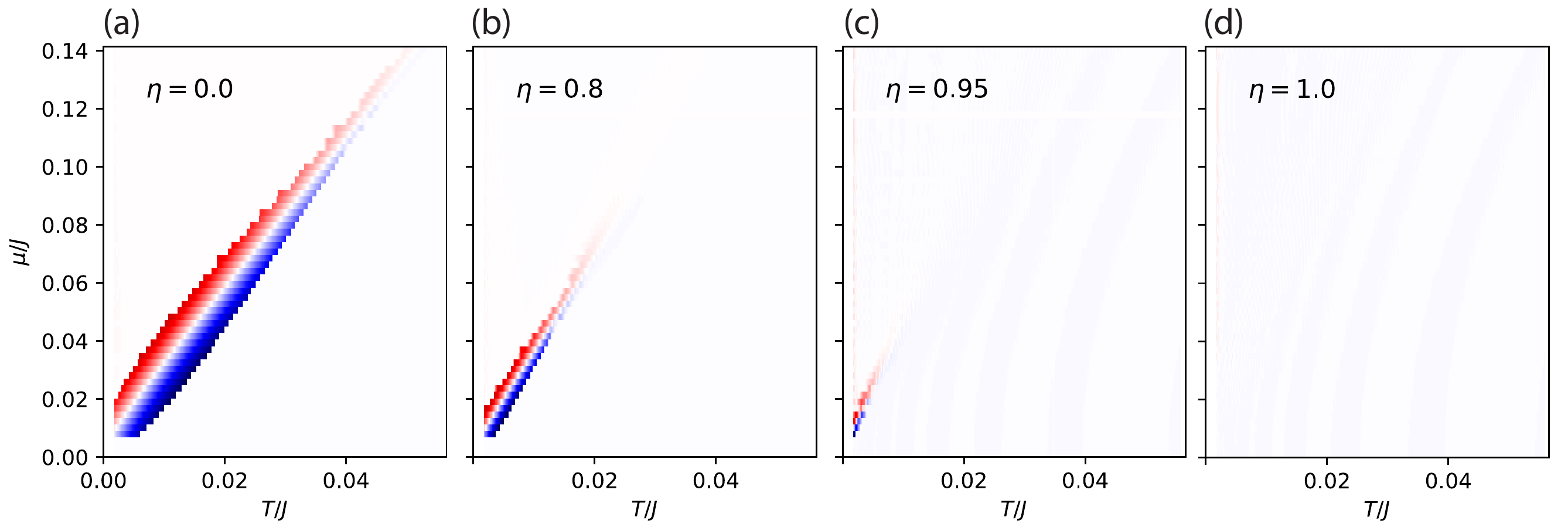}
	\caption{Difference $\Delta{\cal F}$  between the free energy for gapped and gapless phase, indicated by the color for different values of $\eta$. Red (blue) indicates positive (negative) value of  $\Delta{\cal F}$ in arbitrary units. The boundary between two phases shrinks and disappears as $\eta\to 1$.} 
	\label{fig:HPtransition}
\end{figure*}

The boundary separating the gapless phase from the gapped phase is shown in the $(\eta,\mu)$ phase diagram Fig.~\ref{fig:largeNgap}(b). For small values of  $\mu$ the gapped phase exhibits an \emph{approximate} conformal symmetry where $G_{\alpha \beta}(\tau)$ can be approximated by reparametrization of the conformal correlator and the gap scales as $E_{\rm gap} \sim \mu^{2/3}J^{1/3}$ for $\eta < 1$. For general $\mu$, the numerically obtained exponential decay of the imaginary time Green's function gives the gap dependence on $\eta$ as shown in Fig.~\ref{fig:largeNgap}(a). The transition between the gapless and the gapped phase along the $\eta$ axis is continuous; the order parameter vanishes smoothly and the free energy smoothly changes across the transition (not shown). This motivates us to fit the gap to a BCS-type gap function. The functional form $a_{1} \mathrm{tanh}(a_{2} \sqrt{{\eta_{c}/\eta}-1})$ is used to fit to the dependence of the gap on $\eta$ and to find the phase boundary in $(\mu,\eta)$-parameter space close to zero temperature. 
%$\eta_{c}(\mu) = \sqrt{1-a \mu^{3/2}}$.\textcolor{blue}{Find a functional dependence of the gap on $\eta$?}

The system free energy ${\cal F}$ is the large-$N$ action given as Eq.~(\ref{action}) in Appendix~\ref{apdx:sd} evaluated for the saddle point solutions and reads
\begin{eqnarray}
-\frac{\cal F}{N}&=&
\ln(2)
+
\frac{1}{2}
\sum_{\omega_n}
\ln \left(\frac{D(i\omega_{n})}{(i\omega_{n})^2}
\right)
\nonumber \\ &&+
\frac{3}{4}
\sum_{\omega_n}^{} 
\left(
\frac{1}{2}\Sigma_{LL}(\omega_n) G_{LL}(\omega_n)\right.
\\
&& \left. +	\frac{1}{2}\Sigma_{RR}(\omega_n) G_{RR}(\omega_n)
-
\Sigma_{LR}(\omega_n) G_{LR}(\omega_n) \nonumber 
\right)
\end{eqnarray}
The free energy difference $\Delta{\cal F}$ between gapped and gapless phases, as obtained by sweeping the temperature $T$ up and down during the self-consistent solution of the saddle point equations  \eqref{eqn:sd}, is displayed in Fig.~\ref{fig:HPtransition}.
For $\eta<1 $ we observe pronounced hysteresis indicating a first-order Hawking-Page type transition \cite{Hawking1983} between the low temperature SYK phase and the gapped \emph{wormhole} phase. The region in the $\mu$-$T$ phase diagram where the two  phases can coexist shrinks as $\eta$ increases only to finally disappear when $\eta\to 1$. 

%\textit{scaling for $\eta=0, \eta=0.5$}

\subsection{Revival dynamics\label{subsec:Revivaldynamics}}

A prominent signature of the wormhole behavior is the revival dynamics \cite{Plugge2020}, manifest as an oscillatory behavior of the transmission amplitude
\begin{equation}
  T_{\alpha\beta}(t) = 2 | G_{\alpha\beta}^{>}(t)|.
\end{equation}
  Here, the greater Green's function $G_{\alpha\beta}^{>}(t)=\frac{1}{N}\theta(t) \sum_i \langle \chi_{i}^{\alpha}(t)\chi_{i}^{\beta}(0) \rangle$ may be interpreted as the amplitude of recovering the excitation $\chi^{\beta}_{i}\ket{GS}$, inserted on side $\beta=L/R$ and probing on side $\alpha$ at a later time $t$, averaged over all modes $i$. Inside the black hole phase, any excitation is rapidly scrambled and the transmission amplitude decays according to the power law $t^{-1/2}$. In the wormhole phase, periodic revivals are expected where $T_{LR}$ and $T_{LL}$ oscillate out of phase \cite{Plugge2020}.

To compute the real-time Green's functions, we analytically continue the imaginary-time saddle-point  equations
(\ref{eqn:sd}). The detailed procedure is given in Appendix \ref{apd:ac}, yielding the following set of equations for retarded propagators and self energies:
\begin{eqnarray}
\label{eqn:sd_real}
G^{\rm ret}_{\alpha\alpha}(\omega) &=&
\frac{\omega -\Sigma^{\rm ret}_{\bar\alpha\bar\alpha}(\omega)}{D(\omega)},
\nonumber
\\
G^{\rm ret}_{LR}(\omega) &=& 
-\frac{i\mu -\Sigma^{\rm ret}_{LR}(\omega)}{D(\omega)},
\\
\Sigma^{\rm ret}_{\alpha\beta}(\omega) &=& -iJ^2 (1+\alpha\eta)(1+\beta\eta) 
\nonumber
\\
&& \times \int_0^\infty dt e^{i(\omega+i\eta)t}
\left[n^+_{\alpha\beta}(t)^3+n^-_{\alpha\beta}(t)^3 \right].
\nonumber
\end{eqnarray}
Here,
\begin{eqnarray*}
	D(\omega) = 
	\left(\omega - \Sigma^{\rm ret}_{RR}(\omega)\right)
	\left(\omega - \Sigma^{\rm ret}_{LL}(\omega)\right)
	+
	\left(i\mu - \Sigma^{\rm ret}_{LR}(\omega)\right)^2
\end{eqnarray*}
and
\begin{equation}
  n^{s}_{\alpha\beta}(t) = \int_{-\infty}^{\infty} d\omega \rho_{\alpha\beta}(\omega) n_F(s
  \omega) e^{-i\omega t}.
\end{equation}

Equations \eqref{eqn:sd_real} are solved by numerical iteration using the fast Fourier transform algorithm to pass between time and frequency domains. The spectral function is obtained as the imaginary part of the mode-averaged retarded Green's function, $\rho _{\alpha\beta}(\omega) = -\frac{1}{\pi} \text{Im} G^{\rm ret} _{\alpha\beta} (\omega)$.  Using the relation
        \begin{equation}
        G^>_{\alpha\beta}(\omega)=2\pi i[n_F(\omega)-1]\rho _{\alpha\beta} (\omega)
        \end{equation}
one can easily extract the greater Green's function, and thus the transmission amplitude $T_{\alpha\beta}(t)$,  from the knowledge of the spectral function.

\begin{figure*}
	\centering
	\includegraphics[width=\textwidth]{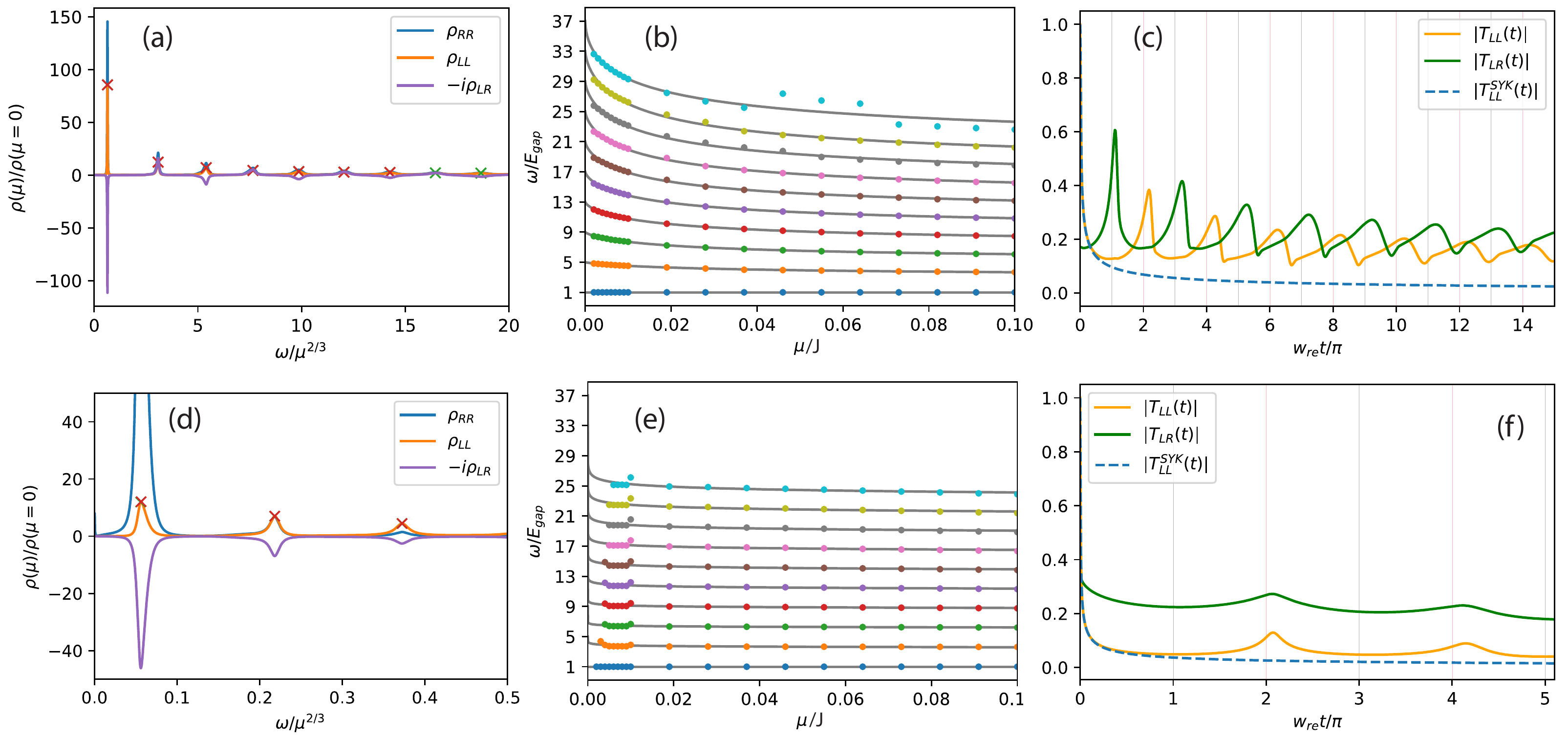}
	\caption{(a,d) Spectral function showing recurring peak structure at different frequencies for $\eta=0.5, 1$. (b,e) Extracted normalized peak position as a function of tunneling strength for $\eta=0.5, 1$. The top plot indicates a conformal tower structure with peak positions at $4n+1$ for small $\mu$ and $\eta=0.5$. Bottom plot for $\eta=1$ does not show conformal tower of the form $4n+1$. (c,f) Transmission amplitude from spectral function for $\eta=0.5, 1$ showing the out-of-phase vs. in-phase relation between diagonal and off-diagonal peaks. (a,c) correspond to $\mu/J=0.004$ and (d,f) correspond to $\mu/J=0.01$.}
	\label{fig:realtime0}
\end{figure*}

At low temperatures, for $0\leq\eta < 1 $ the ground state is gapped as shown before, and additionally the spectral function $\rho(\omega)$ develops a series of equidistant peaks shown in Fig. \ref{fig:realtime0}(a). 
For $\mu/J$ small, they represent the conformal tower of states in the MQ model, $E_n=E_{\rm gap} (4n+1)$ with $n$ integer \cite{maldacenaqi2018,Plugge2020}. 
Here we confirm that this also holds in the imbalanced case as shown in Fig.~\ref{fig:realtime0}(b) for $\eta=0.5$, where we plot the peak energies as a function of $\mu$. Fourier transform of the spectral function  autocorrelation is equal to the absolute square of the transmission amplitude \cite{Plugge2020}. Hence, the equidistant peak structure of $\rho_{LL}$ and $\rho_{LR}$ leads to  revival oscillations in the transmission amplitude of the excitation. Since the peaks in $\rho_{LR}$ have alternating sign structure, there is a $\pi/2$ phase difference between $T_{LL}$ and $T_{LR}$ as shown in Fig.~\ref{fig:realtime0}(c). \\
The large-$N$ approach can be used to indirectly test for the proximity of the ground state to the $\ket{\rm TFD}$ by considering the expectation value $\langle H^{L}_{\rm SYK} - H^{R}_{\rm SYK}\rangle$ which, according to Eq.~\eqref{iden}, should vanish in the TFD state.  As we show in the Appendix~\ref{Appendix:tfdstate},
the gapped ground state obtained by solving the large-$N$ equations \eqref{eqn:sd_real} indeed  has this expectation value very close to zero  for all $0\leq\eta<1$ and non-zero $\mu$. This suggests that the ground state of imbalanced MQ model remains close  to $\ket{\rm TFD}$ for all $N$.

\section{The maximally imbalanced limit}
\label{sec:eta1limit}

\subsection{Wormhole or black hole?}
The saddle point solution at $\eta=1$ shows that the ground state is gapped for finite $\mu$ but the Hawking-Page transition does not occur as temperature increases. This already indicates that the physics for $\eta=1$ is different compared to the ordinary  MQ model. Additional differences are observed in the revival dynamics as indicated in Fig.~\ref{fig:realtime0}(d-f). 
The spectral function still shows a series of peaks but their positions no longer approach the conformal tower values  $E_{\rm gap} (4n+1)$ for small $\mu$. The revival oscillations for the transmission amplitudes, $T_{LL}$ and $T_{LR}$, are now in phase. A related observation is that unlike what we observe for $\eta< 1$, the signs of $\rho_{LR}$ peaks do not alternate. 

Saddle point equations can be used to further analyze the peak structure in the special case $\eta=1$. For $\mu \ll J$, the MQ model ($\eta=0$) has approximate conformal symmetry, and hence the Green's function in the \emph{wormhole} phase have the approximate functional form \cite{maldacenaqi2018} 
\begin{gather}
	G_{\alpha \alpha}(t) = \frac{c}{|2 J \sin(\mathrm{w}_{re} t/2)|^{1/2}} , \\
  G_{LR}(t) = \frac{i c}{|2 J \cos(\mathrm{w}_{re} t/2)|^{1/2}}, \
\end{gather}
where $\mathrm{w}_{re}=4 E_{gap} \sim \mu^{2/3}$, and the distance between the consecutive peaks $\mathrm{w}_{re}(t-t')$ is $2\pi$. This conformal form of the correlator is obtained using time-reparametrization symmetry of the one sided SYK correlator, $G(|t_{1}- t_{2}|) \rightarrow |f'(t_{1})f'(t_{2})|^{\frac{1}{4}} G(|f(t_{1})- f(t_{2})|)$, where the time on the left side transforms as $t \rightarrow -\cot(\mathrm{w}_{re} t/2)$ and on the right side as $t \rightarrow \tan(\mathrm{w}_{re}t/2)$. Note that this results in a $\pi/2$ phase difference between $T_{LL}(t) = G_{LL}(t>0)$ and $T_{LR}(t) = G_{LR}(t>0)$. For other $\eta$, we speculate that the conformal invariance is broken when $\mu/J \ll \sqrt{(1-\eta^2)}$ no longer holds, rendering reparametrization invalid.

\begin{figure}
	\centering
	\includegraphics[width=0.5\textwidth]{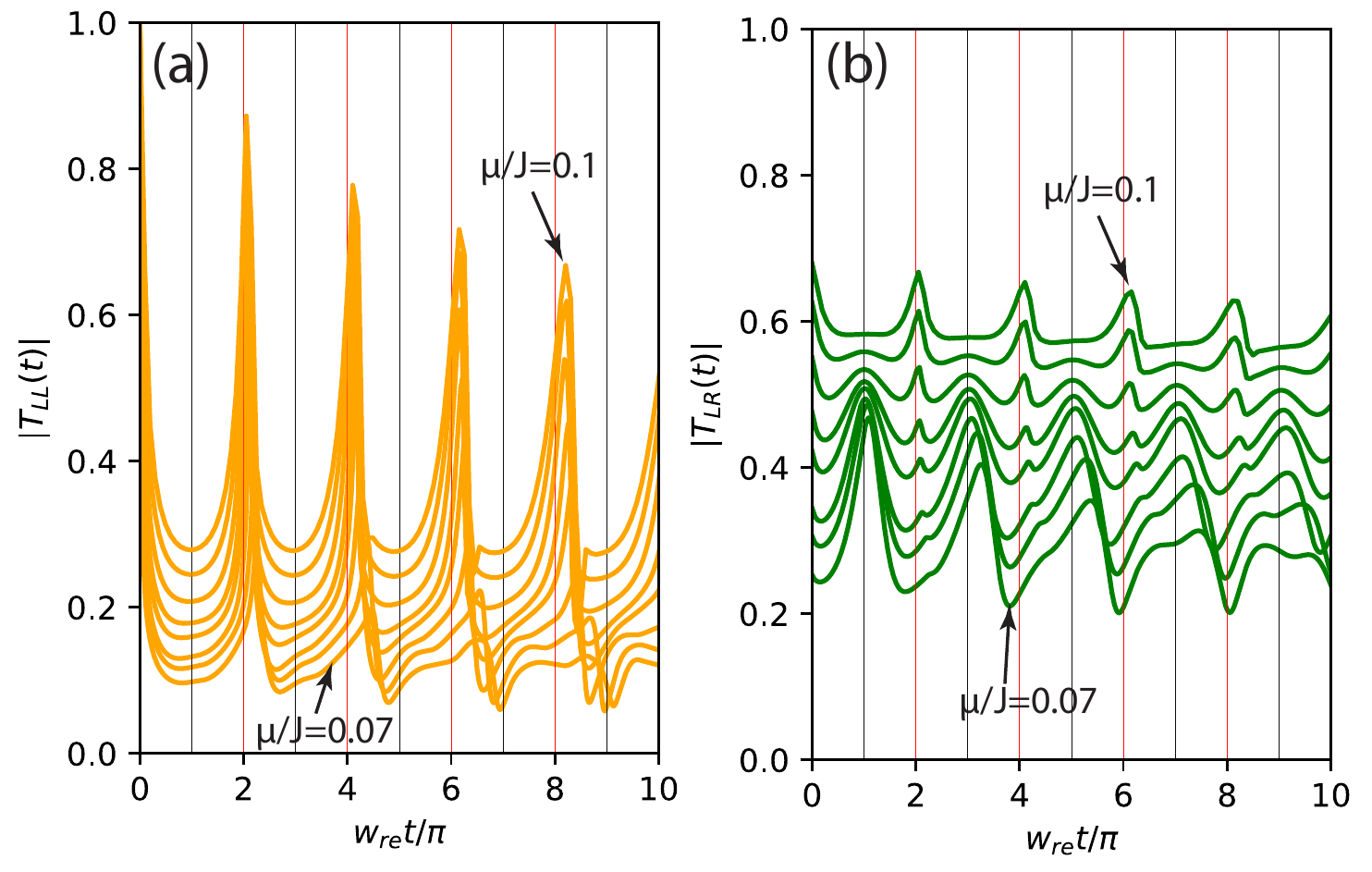}
	\caption{Transmission amplitudes $\left|T_{LL}(t)\right|$ (a) and $\left|T_{LR}(t)\right|$ (b) for $\eta=0.95$ and a range of values  $\mu/J=0.007, 0.01, 0.019, 0.28, 0.037, 0.055, 0.073, 0.1$ from bottom to top. The amplitudes show revival dynamics which evolve from out-of-phase to in-phase with increasing $\mu$.}
	\label{fig:realtimeeta1}
\end{figure}

In Fig.~(\ref{fig:realtimeeta1}) we show, for $\eta$ slightly less than 1, a transition from out-of-phase to in-phase oscillation as $\mu$ increases. At $\eta=1$, we find that the oscillations are in phase for all $\mu$, which can be understood by the following argument. The only non-zero self energy in this case is $\Sigma_{LL}$, thus we can relate diagonal Green's function with off-diagonal one by
\begin{align} 
G_{LL}^{R}(\omega) &= \frac{i\omega}{\mu} G_{LR}^{R}(\omega) \Rightarrow G_{LL}^{R}(t) = \frac{1}{\mu} \frac{\partial}{\partial t} G_{LR}^{R}(t), \nonumber\\ G_{RR}^{R}(\omega) &= \frac{i\omega}{\mu} G_{LR}^{R}(\omega) - \frac{i}{\mu} \Sigma_{LL}^{R}(w) G_{LR}^{R}(\omega).
\end{align}
One can check that taking a numerical derivative of $T_{LR}$ with appropriate multiplicative constant gives us $T_{LL}$ with peaks in phase with $T_{LR}$.  

For $\eta=1$, although the peaks in $T_{LR}$ do not decay sharply, they suggest that the ground state could be a TFD, as also indicated by our ED results. Whether or not it describes a full-fledged wormhole in the gravity context remains an interesting open question which we leave for future study. 

\subsection{Proposed physical realization}

We consider a setup proposed by Chew, Essin and Alicea \cite{Alicea2017} for the physical realization of  the Majorana SYK model. The setup is depicted in Fig.~\ref{fig0}  and consists of an array of proximitized quantum wires weakly coupled to a disordered quantum dot. Wave functions of Majorana zero modes localized on the left end of wires delocalize into the quantum dot and, in the presence of interactions, are described by the SYK model in Eq.~\eqref{eqn:SYK}.  In Ref.~\onlinecite{Alicea2017} very long wires are assumed such that zero modes localized at the right end of wires do not play any role. Here we consider a case with short wires such that $\chi_j^L$ and $\chi_j^R$ are weakly coupled thus furnishing the $H^{LR}$ term required to construct the imbalanced MQ model.  One expects that $\mu \sim e^{-2L/\xi} \cos(k_{F}L)$, with $\xi$ the effective coherence length \cite{Sarma2012}. Intuitively one also  expects that when the $R$ Majorana modes are not coupled to a quantum dot this setup will realize the maximally imbalanced MQ model.
Below we examine the mapping of Ref.~\onlinecite{Alicea2017} to the SYK model more closely and  confirm the validity of this intuition in the limit of short wires.

We first consider the non-interacting Hamiltonian $H_{0}$ describing the setup in Fig.~\ref{fig0} with the $R$ quantum dot absent. According to Ref.~\onlinecite{Alicea2017}
the relevant model  belongs to the BDI symmetry class \cite{Kitaev2009,Ryu2010}, where time-reversal is preserved, and is described by the Hamiltonian
\begin{align}
	H_{0} = \frac{i}{4} 
	\begin{bmatrix} 
	\Gamma^{T} & \tilde{\Gamma}^{T} 
	\end{bmatrix} 
	\begin{bmatrix} 
	0 & M \\ -M^{T} & 0 
	\end{bmatrix} 
	\begin{bmatrix} 
	\Gamma \\ \tilde{\Gamma} 
	\end{bmatrix}.
\end{align}
Here $\Gamma = \left( \eta_{1} \eta_{2} \cdots \eta_{N_{dot}} \; \chi^{L}_{1} \chi^{L}_{2} \cdots \chi^{L}_{N}\right)^{T}$ and $\tilde{\Gamma} = \left( \tilde{\eta}_{1} \tilde{\eta}_{2} \cdots \tilde{\eta}_{N_{dot}}\right)^{T}$ are expressed in terms of 
the dot complex fermions denoted by $\eta_{i}+i\tilde{\eta}_{i}$ and Majorana zero modes $\chi^{L}_{i}$. The next step is to include the coupling to $R$ Majorana fermions. To this end we project the hybridization $H^{LR}$  onto the manifold of the lowest energy eigenstates of $H_{0}$. A singular value decomposition of the rectangular matrix $M$ is performed such that the diagonal matrix is $\Lambda := \mathcal{O}^{T} M \tilde{\mathcal{O}}$, and the resulting new Majorana zero modes are the states corresponding to zero singular values. The original fermion basis is related to the new basis through an orthogonal transformation
\begin{align} 
\Gamma = 
\begin{bmatrix} 
\eta \\ \chi^{L} 
\end{bmatrix} 
= 
\begin{bmatrix} 
\mathcal{O}_{\eta, \eta'} & \mathcal{O}_{\eta, \chi'}  \\ \mathcal{O}_{\chi, \eta'} & \mathcal{O}_{\chi, \chi'} 
\end{bmatrix} 
\begin{bmatrix} 
\eta' \\ \chi'^{L}
\end{bmatrix}.
\end{align} 
where $\chi'^{L} = \left(\chi'^{L}_{1}, \cdots \chi'^{L}_{N}\right)$ are the new $N$ left zero modes. The $LR$ hybridization term in this basis becomes
\begin{align}
\label{basischange} 
H^{LR} = &  i \mu \left(\eta'^{T} \mathcal{O}_{\chi, \eta'}^{T}  \chi^{R} + (\chi'^{L})^{T} \mathcal{O}_{\chi, \chi'}^{T}  \chi^{R}\right) \nonumber\\ 
  \approx & i \mu (\chi'^{L})^{T} \mathcal{O}_{\chi, \chi'}^{T}  \chi^{R}.                                                                                                                                            \end{align}
Since the dot energy level separations are much larger than the tunneling strength $\mu$, we make a reasonable approximation that the first term that hybridizes dot fermions with the right zero modes can be ignored.

                                                                                                                                                                        If the matrix $\mathcal{O}_{\chi, \chi'}$ was orthogonal, then by defining $\chi'^{R} = \mathcal{O}_{\chi, \chi'}^{T}  \chi^{R}$, we would have the same one-to-one tunneling between the primed Majorana modes. This does not apply trivially because the sub-matrix $\mathcal{O}_{\chi,\chi'}$ should be first appropriately normalized to be a $N$ dimensional orthogonal matrix, and even then it is only \emph{approximately}  orthogonal. This can be quantified by writing  $\mathcal{O}_{\chi, \chi'}$  as a sum of its nearest orthogonal matrix $O_{o}$ and a non-orthogonal correction, such that
\begin{align}
\sqrt{\frac{(N_{dot}+N)}{N}}\mathcal{O}_{\chi, \chi'} = O_{o} (I + t^{T}),
\end{align}
where $O_{o} = UV^{T}$ and $\Sigma = U^{T}\mathcal{O}_{\chi, \chi'}V$ is the singular value decomposition of the sub-matrix. After an orthogonal rotation of the right Majorana zero modes, $\chi'^{R} = \mathcal{O}_{o}^{T}  \chi^{R}$, we arrive at
\begin{align}
  H^{LR} \approx i\mu_{\rm eff} (\chi'^{L})^{T}(I+t) \chi'^{R}, 
\end{align}
with $\mu_{\rm eff} = \mu \sqrt{{N}/(N_{\rm dot}+N)}$, which implies a small all-to-all random tunneling term between the $L$ and $R$ primed fermions in addition to the desired one-to-one tunneling.

Two questions now naturally arise: (i) how large is the undesirable coupling $t$,  and (ii) will it spoil the MQ physics that we would like to implement. To answer the first question  
we assume $M$ to be a random matrix chosen from the Gaussian orthogonal ensemble \cite{Beenakker1997}. We then find $t$ to also be a random matrix with zero mean. The variance of the distribution is inferred by evaluating the Frobenius norm of the matrix which is invariant under any orthogonal transformation
\begin{align} 
\|t \|\equiv {\rm Tr}(t^{T}t) = N^2 \langle t_{ij}^2\rangle =  \|I-\Sigma\| \approx 0.2 N, 
\end{align}
where $\langle t_{ij}^2 \rangle = \frac{1}{N^2}\sum_{\{i,j\}=1}^{N}{t_{ij}^2}$ and  $0.2 N$ is the typical value obtained from the random matrix numerical simulation.

The effect of all-to-all random tunneling term  on the imbalanced MQ model is tractable within the replica approach and we analyze it in Appendix \ref{apdx:sd}. We find that the gap in the energy spectrum survives in the presence of this additional perturbation and only reduces in magnitude, thus indicating that the ground state for this setup is adiabatically connected to the $\eta=1$ ground state found previously in the absence of the perturbation. These considerations lead us to conclude that the setup depicted in Fig.~\ref{fig0} with the $R$ quantum dot completely absent could plausibly be used to realize the maximally imbalanced version of the MQ wormhole model.

\section{Discussion and Outlook\label{sec:conclusion}}
Our exact diagonalization and large-$N$ saddle-point results demonstrate that the thermofield double state defined through the  SYK model eigenstates occurs as the ground state of the imbalanced Maldacena-Qi model.  The latter consists of a pair of SYK models with identical interactions up to an overall scale factor. We showed that the gap, the structure of the excitations, and the revival dynamics of two-sided retarded Green's functions all persist as the interaction strengths in two subsystems  become different. These results strongly suggest that the  imbalanced MQ model remains dual to a traversable wormhole. It would be interesting to find the corresponding Schwarzian action and the reparametrization symmetry when left and right conformal correlators are different. Physical consequences of the time evolution asymmetry of the ground state in the imbalanced case are also worth exploring. 

In the maximally imbalanced limit, i.e.\ when one SYK Hamiltonian completely vanishes, the solutions indicate a mixed picture with some wormhole and some black hole characteristic features. For a small coupling between left- and right-side fermions, the early time dynamics of the retarded Green's function shows SYK power-law decay, and a weak revival oscillations are observed at later times. The $LL$ and $LR$ oscillations are in phase, which is unlike the canonical MQ model where they occur $\pi/2$ out of phase, consistent with a gravity interpretation of a particle bouncing back and forth between two ends of the wormhole. The ground state is shown to be gapped and, importantly, our results indicate that it is close to the TFD state. This could be useful since the TFD state can be prepared as a ground state with just a single system emulating the SYK model, thus obviating the difficult challenge of preparing two microscopically identical SYK systems. Finally, we have shown how this maximally imbalanced limit could be physically realized in the previously proposed setup, thus making our analysis relevant for future experiments seeking to produce the elusive TFD state in a solid-state setting.

We conclude by noting that the adiabaticity argument we employed here is completely general and applies to any model where the ground state of two coupled subsystems is close to TFD.

	\begin{acknowledgments}
		We would like to thank Stephan Plugge and \'{E}tienne Lantagne-Hurtubise for helpful discussions and GoogleX for hosting the workshop ``Quantum gravity in the lab'' where the basic idea behind this work was originally conceived. This work was supported by
NSERC, the Max Planck-UBC-UTokyo Centre for Quantum Materials and the Canada First Research Excellence Fund, Quantum Materials and Future Technologies Program.  TH's research at Perimeter Institute
is supported in part by the Government of Canada
through the Department of Innovation, Science and Industry Canada and by the Province of Ontario through
the Ministry of Colleges and Universities.
	\end{acknowledgments}
	\bibliography{biblio}
	\onecolumngrid
	
	\appendix
	
	\section{Derivation of the saddle point equations}
	\label{apdx:sd}

      The imaginary-time path integral defining the  partition function is
\begin{equation}
\label{eq:partition}
\mathcal{Z}=\int \mathcal{D}\chi e^{-
	\int_{0}^{\beta} d\tau \left[\sum\limits_{i,\alpha}^{}   \frac{1}{2}\chi_i^\alpha(\tau)
	\partial_\tau \chi_i^\alpha(\tau)
	+
	H(\tau)\right]}
\end{equation}
where $H$ is given in Eq.~\eqref{modelH}.   	We perform a quenched average over random couplings $J_{ijkl}$ using the replica trick. The variance for the random distribution is $\sigma^2 = {6 J^2}/{N^3}$ with integration measure $\mathcal{D}J_{ijkl} = \prod_{i<j<k<l} dJ_{ijkl}$, where the product is over all independent couplings. The partition function in Eq.\eqref{eq:partition} can be written using Gaussian integral
	\begin{eqnarray}
		\frac{1}{n}\overline{\mathcal{Z}^{n}} =
		\int_{}^{}\mathcal{D}\chi\int_{}^{}\mathcal{D}J_{ijkl}
		\exp \left[- \frac{J_{ijkl}^2}{2 \sigma^2}\right]
		\underbrace{
			\exp \left[-(S_{0}^L +S_{0}^R+
			H^{LR})\right]
		}_{\mathcal{Z}_0}
		\exp \left[
		-\sum_{\alpha}(1+\alpha\eta)\int_{0}^{\beta}d\tau J_{ijkl}
		\chi_{i}^\alpha\chi_{j}^\alpha\chi_{k}^\alpha\chi_{l}^\alpha
		\right],
	\end{eqnarray}
with  $\mathcal{D}\chi= \prod\limits_{i=1}^{N}(d\chi_{ia}^{L} d\chi_{ia}^{R})$ and $\mathcal{Z}_{0}$  the non-interacting part of the action. We made the usual assumption that the solution of interest is diagonal in the replica indices. The averaged partition function is thus
	\begin{eqnarray}
		\mathcal{Z} 
		&=&
		\int_{}^{}\mathcal{D}\chi
		\ {\mathcal{Z}_0}\exp \left[
		\frac{J^2}{8N^3}\sum_{\alpha\beta}^{}
		(1+\alpha\eta)(1+\beta\eta)
		\int_{0}^{\beta}d\tau d\tau'
		\left(\sum_{i}^{}\chi_{i}^\alpha(\tau)\chi_{i}^\beta(\tau')\right)^4
		\right]
	\end{eqnarray}
	Introducing averaged Green's function $G_{\alpha\beta}(\tau,\tau') =
		\frac{1}{N}\mathcal{T}\braket{\sum_{i}^{}\chi_{i}^\alpha(\tau)\chi_{i}^\beta(\tau')}$ using self energies $\Sigma_{\alpha\beta}(\tau,\tau')$ as Lagrange multipliers, we have following form for the action $S$,
	\begin{eqnarray}
		S=&-&\int_{}^{}\mathcal{D}\chi 
		\int_{}^{}\mathcal{D}G\mathcal{D}\Sigma 
		\ \exp \left[
		-\frac{1}{2}\sum_{\alpha\beta}^{} \int_{0}^{\beta} d\tau d\tau'
		\left\{
		\sum_{i}^{}
		\chi_{i}^\alpha(\tau)
		\left(
		\delta^{\alpha\beta}\delta(\tau-\tau')\partial_\tau
		-\Sigma_{\alpha\beta}(\tau,\tau')
		\right)
		\chi_{i}^\beta(\tau')
		\right\}
		\right] \nonumber
		\\
		&\times&
		\exp \left[
		-\sum_{\alpha\beta}^{} \int_{0}^{\beta} d\tau d\tau'
		\left\{
		\frac{N}{2}\Sigma_{\alpha\beta}(\tau,\tau')
		G_{\alpha\beta}(\tau,\tau')
		-N\frac{J^2}{8}(1+\alpha\eta)(1+\beta\eta) 
		G_{\alpha\beta}(\tau,\tau')^4
		\right\}
		\right]
		\\
		&\times&
		\exp \left[
		-i \int_{0}^{\beta} d\tau
		\frac{\mu}{2}N \left(G_{LR}(\tau,\tau) -
		G_{RL}(\tau,\tau) \right) 
		\right].\nonumber
	\end{eqnarray}
	Note that this is exactly the Maldacena-Qi action \cite{Maldacena2018} except for the  $\eta$-dependent prefactor on the second line. Time-translation symmetry, $G(\tau, \tau') = G(\tau-\tau')$, can be used to further simplify the action. The quadratic functional integral over the Fermi fields is performed in the frequency space where the non-interacting part of the action becomes diagonal. Using the following convention for Fourier transforms
	\begin{align}
	G(\tau) = \frac{1}{\beta} \sum_{\omega_{n}} e^{-i\omega_{n}\tau} G(\omega_{n}), \quad G(\omega_{n}) = \int_{0}^{\beta} d\tau e^{i\omega_{n}\tau} G(\tau)
	\end{align}
	with Matsubara frequencies $\omega_{n} = {(2n+1)\pi}/{\beta}$,  the effective action becomes
	\begin{eqnarray}\label{action}
		-\frac{S}{N}&=&
		\frac{1}{2}
		\sum_{\omega_n}
		\ln \det 
		\begin{pmatrix}
			i\omega_n 
			-\Sigma_{RR}(\omega_n)	
			&
			-i\mu 
			- \Sigma_{RL}(\omega_n)
			\\
			i\mu 
			- \Sigma_{LR}(\omega_n)
			&
			i\omega_n
			- \Sigma_{LL}(\omega_n)
		\end{pmatrix}
		\\
		&-&
		\frac{1}{2}\sum_{\alpha\beta}^{}
		\left\{
		\sum_{\omega_n}^{} 
		\left(
		-\Sigma_{\alpha\beta}(\omega_n)
		G_{\beta\alpha}(\omega_n)
		\right)
		-\frac{J^2}{4}(1+\alpha\eta)(1+\beta\eta) 
		\int_{0}^{\beta} d\tau d\tau'
		G_{\alpha\beta}(\tau,\tau')^4 
		\right\}. \nonumber 
	\end{eqnarray}	
	The saddle-point equations in Eq.~\eqref{eqn:sd} of the main text are obtained by minimizing the action, taking  ${\delta S}/{\delta G} = 0$, and  ${\delta S}/{\delta \Sigma} = 0$.

        \subsection*{Large-$N$ equations for disordered tunneling}
\begin{figure}[h]
	    \centering
	    \includegraphics[width=0.5\columnwidth]{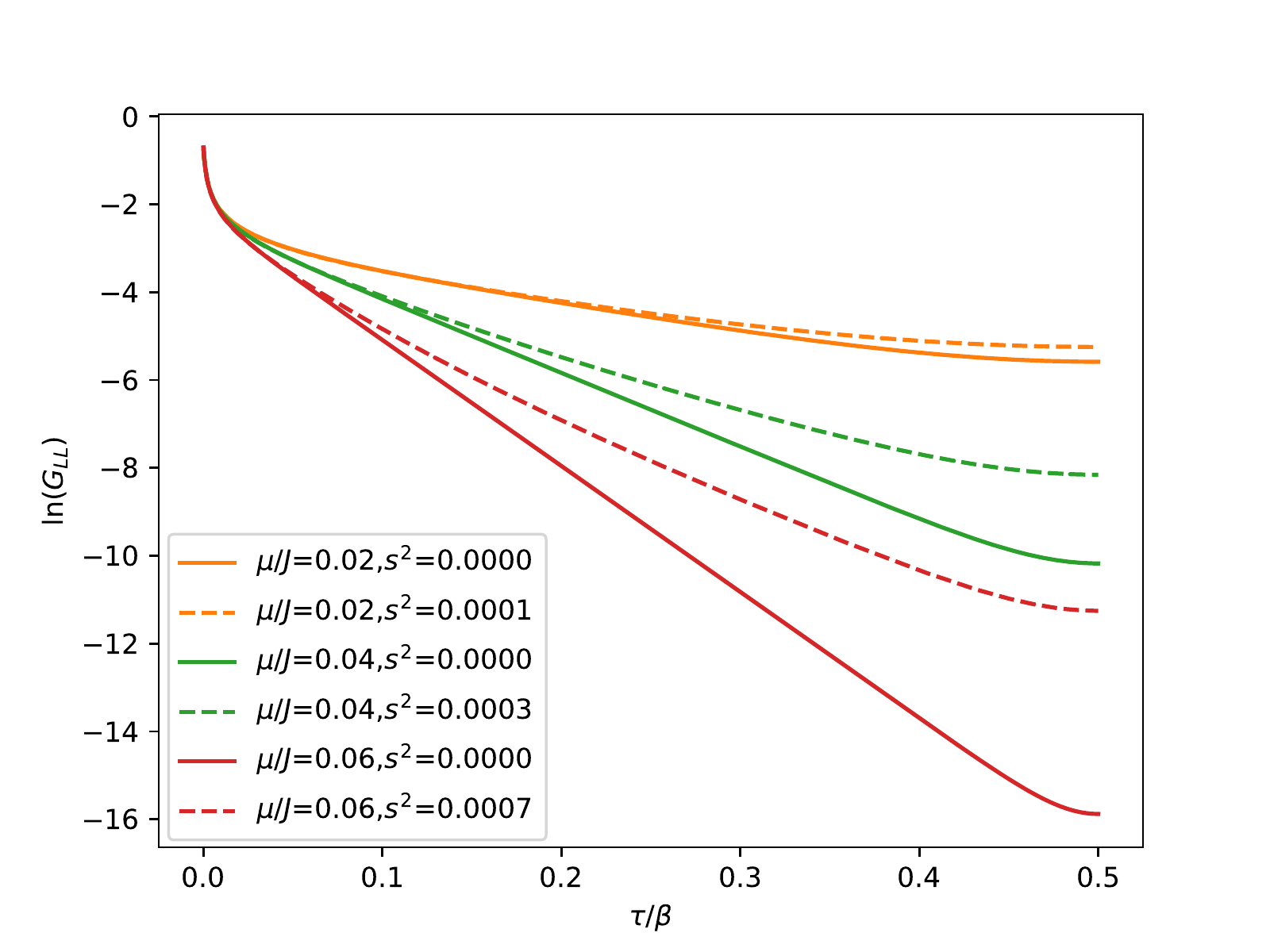}
	    \caption{Semi log plot of $G_{LL}$ for $\eta=1$ showing exponential decay for three different values of $\mu/J$ with and without the random tunneling term. Taking $s^{2} \approx 0.2\mu^{2}$ we find that the gap doesn't close albeit exhibiting a small reduction in the presence of intra-wire Majorana hybridization.}
	    \label{fig:eta1correlator}
	\end{figure}

	As discussed in Sec.~\ref{sec:eta1limit}, when Majorana zero modes hybridize within the wire, after projecting onto states in the disordered quantum dot, the partition function has an extra all-to-all tunneling. The non-interacting part of the Hamiltonian is modified with a random tunneling part and reads
        \begin{align}
          H^{LR} = i\mu_{\rm eff} \left(\sum_{i=1}^{N} \chi^{L}_{i} \chi^{R}_{i} + \sum_{\{i,j\}=1}^{N} t_{ij} \chi^{L}_{i} \chi^{R}_{j}\right),
        \end{align}
 where 
 $\langle (\mu  t)_{ij}^2\rangle = s^2/N\simeq (0.2\mu_{\rm eff}^2)/N$ and $\mu_{\rm eff} = \mu \sqrt{{N}/{(N_{\rm dot}+N)}}$. The $t_{ij}$ are independent Gaussian random variables and we average over them in the same way we did for $J_{ijkl}$ to get get the modified partition function
 \begin{align}
   \mathcal{Z}' = \mathcal{Z} \  \overline{e^{-i\mu_{\rm eff}\sum_{ij}\int d\tau t_{ij}\chi^{L}_{i}(\tau)\chi^{R}_{j}(\tau)}} = \mathcal{Z} \  e^{\frac{s^{2}}{2} \int d\tau d\tau' G_{LL}(\tau,\tau')G_{RR}(\tau,\tau')}
 \end{align}
 Deriving the new set of saddle point equations we find that self energies get modified as follows, $ \Sigma_{LL}' = \Sigma_{LL} + s^{2} \ G_{RR},$ and  $\Sigma_{RR}' = \Sigma_{RR} + s^{2} \ G_{LL}$. The solution of the modified equations is plotted in Fig.~\ref{fig:eta1correlator} and  demonstrates that the gap does not close. This suggests that the ground state of the Chew-Essin-Alicea setup with intra-wire Majorana hybridization is adiabatically connected to the TFD state.

\section{Analytical continuation of the Saddle Point Equations}
\label{apd:ac}
The first two of Eqs.~\eqref{eqn:sd} are transformed to real-time representation by the usual prescription that connects retarded and Matsubara propagators, namely $G^{\rm ret}(\omega)=G(i\omega_n\to \omega+i\delta)$ with $\delta$ a positive infinitesimal. 
The equation for the self energy requires more effort as we want to recast it in a form that will be suitable for iterative numerical solution. To this end we employ the procedure originally devised by Banerjee and Altman \cite{Altman2016} and subsequently used in various related contexts \cite{LH2019,Plugge2020,sahoo2020}. 

For the sake of simplicity we consider the self energy equation $\Sigma(\tau) = G(\tau)^3$
where, compared to Eq.~\eqref{eqn:sd}, we have temporarily suppressed all prefactors and focused on a single component of the $2\times 2$ matrix equation.  After Fourier transforming we obtain  
	\begin{eqnarray}
		\Sigma(i\omega_n)
		=
		\frac{1}{\beta^2} \sum_{n_1 n_2}^{} G(i\omega_{n_1})
		G(i\omega_{n_2}) G \left(
			i\omega_n - i\omega_{n_1} - i\omega_{n_2}
			\right).  
	\end{eqnarray}
We express each Green's function in its spectral representation 
	$G(i\omega_k) = \int_{-\infty}^{\infty}d\omega {\rho(\omega)}/(i\omega_k-\omega)$, where $\rho(\omega) = -\frac{1}{\pi} \text{Im} G^{\rm ret}(\omega)$ is the spectral function and write
	\begin{eqnarray*}
		\Sigma(i\omega_n) = \int_{-\infty}^{\infty} d\omega_1 d\omega_2
		d\omega_3
		\rho(\omega_1)
		\rho(\omega_2)
		\rho(\omega_3)
		\frac{1}{\beta^2}
		\sum_{n_1 n_2}^{}
		\frac{1}{i\omega_{n_1}-\omega_1}
		\frac{1}{i\omega_{n_2}-\omega_2}
		\frac{1}{i(\omega_{n}-\omega_{n_1}-\omega_{n_2})-\omega_3}
	\end{eqnarray*}
        We next perform the two Matsubara sums using standard techniques. Details of this calculation can be found e.g.\ in Appendix D of Ref.~\onlinecite{sahoo2020} and the result is
	\begin{eqnarray}\label{B2}
		\Sigma(i\omega_n) = \int_{-\infty}^{\infty} d\omega_1 d\omega_2
		d\omega_3
		\rho(\omega_1)
		\rho(\omega_2)
		\rho(\omega_3)
		\frac{
			n_F(\omega_1)n_F(\omega_2)n_F(\omega_3) 
			+n_F(-\omega_1)n_F(-\omega_2) n_F(-\omega_3)
		}
		{i\omega_n-(\omega_1+\omega_2+\omega_3)}.
	\end{eqnarray}
We analytically continue this expression, $i\omega_n \rightarrow \omega+i\delta$, to obtain the retarded self energy
	\begin{eqnarray}
		\Sigma^{\rm ret}(\omega) &=& \int_{-\infty}^{\infty} d\omega_1 d\omega_2
		d\omega_3
		\rho(\omega_1)
		\rho(\omega_2)
		\rho(\omega_3)
		\frac{
			P(\omega_1,\omega_2,\omega_3)
		}
		{\omega-(\omega_1+\omega_2+\omega_3)+i\delta}
		\\
		&=&
		-i\int_{0}^{\infty}dt\int_{-\infty}^{\infty} d\omega_1 d\omega_2
		d\omega_3
		e^{i(\omega+i\delta-\omega_1-\omega_2-\omega_3)t}
		\rho(\omega_1)
		\rho(\omega_2)
		\rho(\omega_3)
		P(\omega_1,\omega_2,\omega_3),\nonumber
	\end{eqnarray}
where $P(\omega_1,\omega_2,\omega_3)$ denotes the numerator in Eq.~\eqref{B2}.
Finally, defining 
	\begin{eqnarray}
		n^{s}(t) = \int_{-\infty}^{\infty} d\omega \rho(\omega) n_F(s
		\omega) e^{-i\omega t},
	\end{eqnarray}
with $s=\pm$,	we can rewrite the retarded self energy in a more compact form as
	\begin{eqnarray}
		\Sigma^{\rm ret}(\omega) = -i \int_{0}^{\infty} dt e^{i(\omega+i\delta)t}
		\left[ 
		(n^+(t))^3
		+(n^-(t))^3
		\right].
	\end{eqnarray}
        Restoring the prefactors and matrix indices leads to the last expression in Eq.~\eqref{eqn:sd_real}.

        The full set of equations \eqref{eqn:sd_real} can be used to solve for real-time Green's functions in a way that requires only algebraic manipulations and Fourier transforms which can be performed efficiently using the fast Fourier transform algorithm.

\section{Proximity of the ground state to $\ket{\rm TFD}$}
\label{Appendix:tfdstate}
\begin{figure}[h]
	\centering
	\includegraphics[width=0.7\columnwidth]{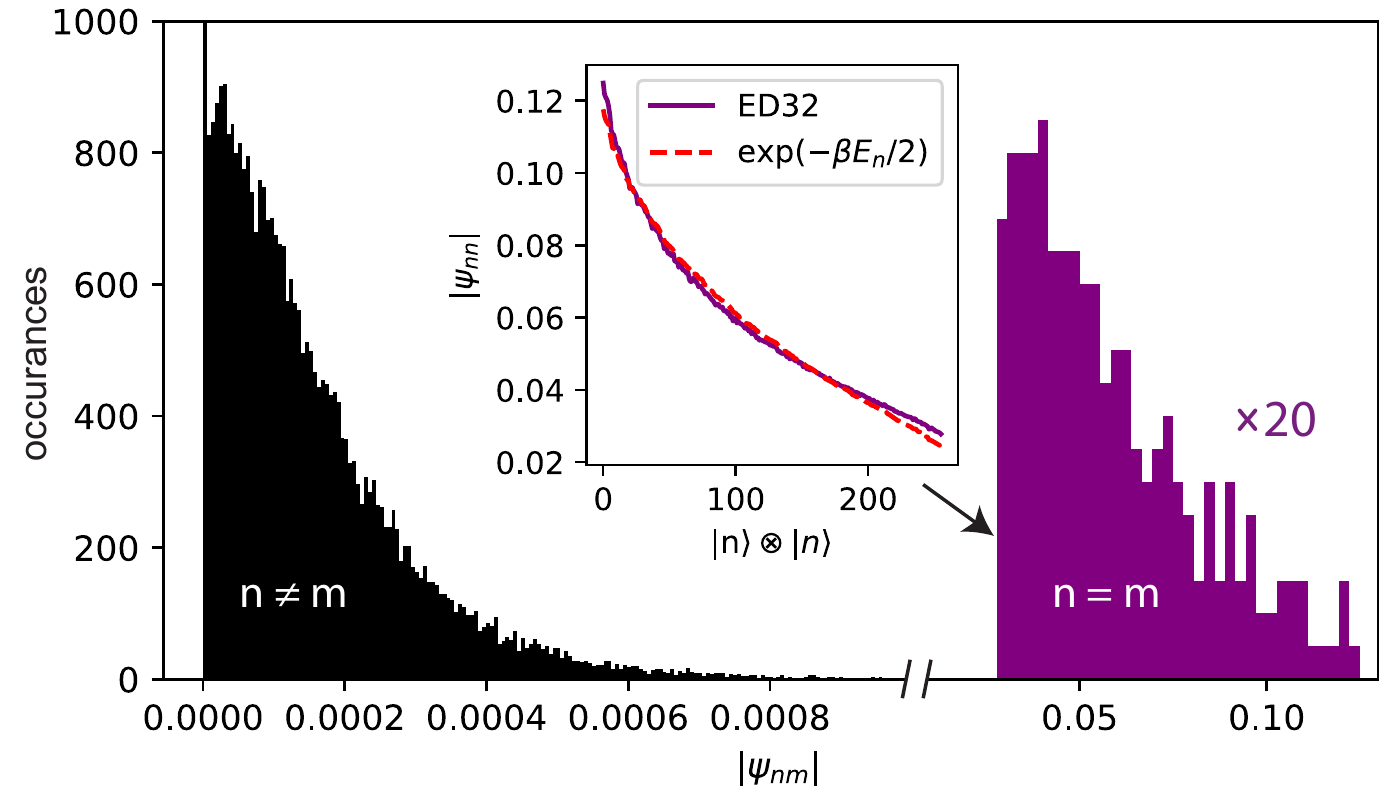}
	\caption{Histogram of absolute values of coefficients $\left|\psi_{nm}\right|$ in the decomposition of  the ground state $\ket{GS}=\sum_{nm}\psi_{nm} \ket{n} \otimes \ket{m}$, obtained by exact diagonalization for $2N=32$ Majoranas. Diagonal values (purple) are non-zero and approximately follow the Gibbs distribution $\exp(-\beta E_n/2)$ (inset). Off-diagonal terms acquire small values distributed around zero (black).}
	\label{fig:coefficients}
\end{figure}
Let us finally discuss the dependence of the ground state wave function $\ket{GS}$ on $\eta$. 
The actual ground state of the imbalanced MQ model can be expressed in terms of the uncoupled SYK basis 
$\ket{GS}=\sum_{nm} \psi_{nm} \ket{n}\otimes \ket{m}$. 
Deviations of $\ket{GS}$ from the TFD state in Eq.~\eqref{eqn:TFD} can occur in two ways: 
(i)  either the diagonal coefficients $\psi_{nn}$ do not exactly follow the
Gibbs distribution $e^{-\tilde{\beta}E_n/2}$, or 
(ii) off-diagonal coefficients become non-zero, i.e.\ $\psi_{nm}\ne \delta_{nm} \psi_{nn}$.
In the latter case we find that $(H^L-H^R)\ket{GS}\ne 0$ which would result in increasing deviations of the ground state from $\ket{\rm TFD}$ as $\eta$ grows. Our ED calculations reveal that for $2N=16$
off-diagonal terms are exactly zero up to machine precision. Here, the fidelity is less than $1$ only for reason (i). For $2N=32$, off-diagonal amplitudes become finite as shown in Fig.~\ref{fig:coefficients}.
The exact TFD state satisfies $(H^{L}_{\rm SYK}-H^{R}_{\rm SYK}) |\mathrm{TFD}\rangle = 0$. In Fig.~\ref{fig:coefficients}, the numerical ground state for $\eta=0$ shows deviations from the TFD state for $2N=32$, indicating that the above condition is only approximately satisfied.%

A weaker condition that verifies if the ground state is a TFD in the large-$N$ limit is to check if the ground state expectation value $\braket{H^{L}_{\rm SYK}-H^{R}_{\rm SYK}} = \braket{\frac{1}{(1+\eta)}H^{L}-\frac{1}{(1-\eta)}H^{R}}$ vanishes, as expected on the basis of Eq.~\eqref{iden}. To evaluate this expression we take advantage of the fact that the expectation value of the system Hamiltonian can be computed from the Green's function. Specifically, for the MQ model one can show that the following expression holds \cite{LH2019}
\begin{align}
  \langle H^{\alpha}\rangle = \frac{N}{4}\lim_{\tau' \rightarrow \tau^{+}} \left[\partial_{\tau} G_{\alpha \alpha}(\tau'-\tau) + i\mu G_{LR}(\tau'-\tau) \right],
\end{align}
for $\alpha=L,R$.  Passing to the Matsubara frequency domain we find
	\begin{align}
	\langle H^{L}_{\rm SYK}-H^{R}_{\rm SYK} \rangle= \frac{N}{4\beta}\sum_{\omega_{n}}e^{-i\omega_{n}0^{+}}\left[\frac{i\omega_{n} G_{LL}(i\omega_{n}) - i\mu G_{LR}(i\omega_{n})}{1+\eta} - \frac{i\omega_{n} G_{RR}(i\omega_{n}) - i\mu G_{LR}(i\omega_{n})}{1-\eta}\right] 
	\end{align}
        We next use the spectral representation, $G_{\alpha\beta}(i\omega_{n}) = \int_{-\infty}^{\infty} d\omega \frac{\rho_{\alpha\beta}(\omega)}{i\omega_{n}-\omega}$, such that $\rho_{LL} = -\frac{1}{\pi} \mathrm{Im}\left[G_{LL}^{\rm ret}(\omega)\right]$ and $\rho_{LR} = \frac{i}{\pi} \mathrm{Im}\left[iG_{LR}^{\rm ret}(\omega)\right]$. Summing over Matsubara frequencies we have
        \begin{align} \langle H^{L}_{\rm SYK}-H^{R}_{\rm SYK} \rangle=\frac{N}{4}\int_{-\infty}^{\infty} d\omega  n_{F}(\omega) \left[\omega \left(\frac{\rho_{LL}(\omega)}{1+\eta}-\frac{\rho_{RR}(\omega)}{1-\eta}\right) - \frac{2 \mu \eta}{1-\eta^{2}} i\rho_{LR}(\omega)\right] ,
        \end{align}
        where we used
        \begin{align}\frac{1}{\beta} \sum_{i\omega_{n}}\frac{e^{i\omega_{n}0^{+}}}{i\omega_{n}-\omega} = n_{F}(\omega), \quad \frac{1}{\beta} \sum_{i\omega_{n}}\frac{i\omega_{n}e^{i\omega_{n}0^{+}}}{i\omega_{n}-\omega} = \omega n_{F}(\omega).
                      \end{align}
                      
\begin{figure}[H]
	\centering
	\includegraphics[width=0.5\columnwidth]{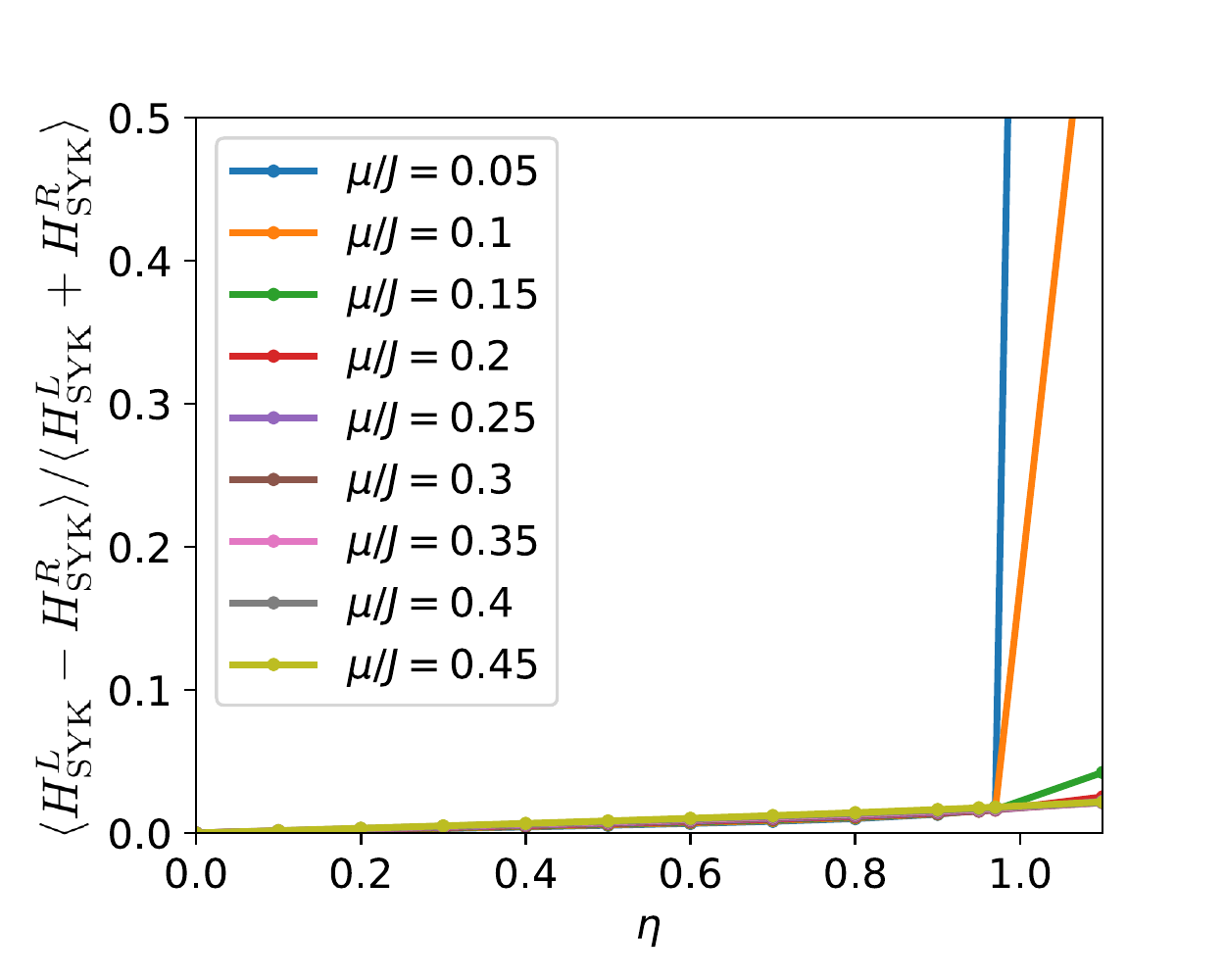}
	\caption{SYK Hamiltonian expectation value ratio computed from the large-$N$ solution as a function of $\eta$ for $T/J=0.001$. The value remains small compared to 1 for any $\eta<1$.}
	\label{fig:HLminusHR}
\end{figure}

 In Fig.~\ref{fig:HLminusHR}, we plot a ratio, $\langle H^{L}_{\rm SYK} - H^{R}_{\rm SYK}\rangle/\langle H^{L}_{\rm SYK} + H^{R}_{\rm SYK}\rangle$, that should vanish for an exact TFD. Note that for $\eta=0$, this is exactly zero by the symmetry of Green's function where $G_{LL} = G_{RR}$, but this symmetry argument does not work for non-zero $\eta$. If the ground state is close to a TFD we expect the ratio to be close to zero. This is indeed what Fig.~\ref{fig:HLminusHR} shows for all values of $\mu$ and $\eta$, except for smallest $\mu$ in the close vicinity of $\eta=1$, in accord with our finding that the system becomes gapless in this limit.
\end{document}